# Towards Augmenting Tip-Enhanced Nanoscopy with Optically Resolved Scanning Probe Tips

Jeremy Belhassen[1,4], Simha Glass[1,2], George A. Stanciu[3], Denis E. Tranca[3], Zeev Zalevsky[3,4*], Stefan G. Stanciu,[3,*] Avi Karsenty[1,2*]

[1]Adv. Lab. of Electro-Optics (ALEO), Faculty of Eng., Lev Academic Center, 9116001, Jerusalem, Israel

[2]Nanotechnology Center for Education and Research, Lev Academic Center, 9116001, Jerusalem, Israel

[3]Center for Microscopy-Microanalysis and Information Processing, Politehnica University of Bucharest, Romania

[4]Faculty of Engineering, Bar-Ilan University, Ramat Gan, 5290002, Israel

[5]The Nanotechnology Center, Bar-Ilan University, Ramat Gan, 5290002, Israel

Corresponding authors: zeev.zalevsky@biu.ac.il ; stefan.g.stanciu@upb.ro; karsenty@g.jct.ac.il

**Abstract:**

A thorough understanding of biological species and of emerging nanomaterials requires, among others, their in-depth characterization with optical techniques capable of nano-resolution. Nanoscopy techniques based on tip-enhanced optical effects have gained over the past years tremendous interest given their potential to probe various optical properties with resolutions depending on the size of a sharp probe interacting with focused light, irrespective of the illumination wavelength. Although their popularity and number of applications is rising, tip-enhanced nanoscopy techniques (TEN) still largely rely on probes that are not specifically developed for such applications, but for Atomic Force Microscopy. This cages their potential in many regards, e.g. in terms of signal-to-noise ratio, attainable image quality, or extent of applications. In this article we place first steps towards next-gen TEN, demonstrating the fabrication and modelling of specialized TEN probes with known optical properties. The proposed framework is highly flexible and can be easily adjusted to be of o benefit to various types of TEN techniques, for which probes with known optical properties could potentially enable faster and more accurate imaging via different routes, such as direct signal enhancement or novel signal modulation strategies. We consider that the reported development can pave the way for a vast number of novel TEN imaging protocols and applications, given the many advantages that it offers.

## 1. Introduction

Chemical and structural imaging with nano resolution under ambient conditions is of utmost importance for advancing our knowledge on biological processes occurring at the sub-cellular level[1-3], which is needed for achieving a thorough understanding of severe diseases such as cancers, neurodegenerative or autoimmune disorders[4]. Nano-resolution is also beneficial for understanding the operando-behaviour of emerging advanced materials[5], or for resolving interfaces in opto-electronic devices, facilitating their exploitation. These perspectives motived over the past decades a consistent body of work in the field of nanoscale imaging, which led to the advent of a wide variety of techniques, each with its own strengths and limitations. Among these, fluorescence-based super-resolution microscopy (f-SRM) techniques succeed in overcoming the resolution limits imposed by diffraction, reaching resolutions in the 20-100 nm range[3,6]. However, they still suffer of important drawbacks: e.g. i) they require very specialized fluorescent probes[7,8], ii) some f-SRM techniques rely on laser beam exposure levels that can lead to phototoxicity and photodamage[9], and iii) recent studies suggest that unpredictable anomalous processes related to the distribution of f-SRM dyes in biological samples exist[10]. Furthermore, due to the requirement of fluorescent labelling their use is severely limited when it comes to resolving physico-chemical properties of advanced nanomaterials, or of nanostructured devices, that cannot be labelled.

Another family of techniques, based on tip enhanced optical effects, has emerged over the past years as a valuable alternative that can successfully overcome the abovementioned limitations of f-SRM. Techniques belonging to this family, such as scattering-type scanning near-field optical microscopy (s-SNOM)[11], tip-enhanced Raman Spectroscopy (TERS)[12], tip-enhanced fluorescence (TEF)[13], Second Harmonic Generation – Scanning Near-Field Optical Microscopy (SHG-SNOM), tip-enhanced photoluminescence (TEPL)[14], or Photo-induced Force Microscopy (PiFM)[15] have gained very high-interest as they can extract optical properties at nanoscale resolution dictated by the size of the tip used for scanning the sample, irrespective of the wavelength used for excitation. Other important features of these techniques are represented by their suitability to operate in ambient conditions and using very low power excitation conditions, not harmful to the investigated samples. Landmark experiments have shown that these techniques can be harnessed to yield resolutions even below 1 nm[14,16]. A major advantage of these techniques is that they share an important overlap in their architecture, given that all rely on the illumination of a sharp, Atomic Force Microscopy-like, probe (a.k.a tip), that scans the surface of the sample, and the subsequent detection of optical signals associated to the light-probe interaction. Such overlap enables facile correlative use, which is highly useful for experiments that require in-tandem access to complementary information that can be provided by such techniques[17-20]. Approaches of these kind have great value considering the rich range of diverse information that can be accessed by Tip-Enhanced Nanoscopy (TEN) techniques relying on distinct principles, as discussed in the next paragraph.

The contrast mechanisms of s-SNOM enable mapping the dielectric properties[21-24] of samples and their chemical composition[11], which have facilitated so far a wide range of discoveries in the condensed phase materials, and two-dimensional materials[25-33]. Additionally, s-SNOM applications in biology enable precise mapping of specific cellular and sub-cellular structures of interest[34-37], and even mapping the exact values of important optical parameters such as the refractive index[38]. TERS, combining the advantages of surface-enhanced Raman spectroscopy and scanning probe microscopy technologies[39-41], found its utilization in the investigation of a wide variety of nanostructured materials ranging from organic to inorganic chemical substances. Among the prominent TERS applications, the characterization of carbon materials such as carbon nanotubes[42-44] and graphene[45] can be easily noticed, as these materials are very stable during measurement and exhibit strong Raman scattering intensities[43,44,46,47]. While carbonaceous materials are probably the most investigated samples by TERS, various biological species have also received great attention[48,49]. Several categories, such as specific molecules[50], viruses[51,52], bacteria[53], erythrocytes[54], membrane receptor ligands[55] or cellular systems[56] have been studied and nanoanalyzed by TERS. SHG-SNOM relies on enhancements of the second-harmonic generation (SHG) of either the tip or the sample[57,58], and represents a superb tool for probing nonlinear optical properties of nanostructured surfaces or surface plasmon interactions (and their modifications) which are of importance in the context of surface-enhanced nonlinear phenomena. Among others, SHG-SNOM has been found useful for local imaging and study of ferromagnetic and ferroelectric domains[59]. Another technique that relies on the tip-light interactions, that has brought significant added value to the TEN field is PiFM[15], which has been demonstrated as a powerful tool for probing the local polarizability of a sample at nanoscale resolutions. PiFM was shown to be highly useful for probing linear and nonlinear[60] optical properties of materials that exhibit spatial variations on the nanoscale[61] and detectable spectroscopic contrast[62]. Specific applications such as visualizing the electric field distributions associated with propagating surface-plasmon-polaritons[63], or identification of contaminants and defects in nanostructured materials[15] hint that PiFM will play a key role in the coming years in material and device characterization. Although TEF, lacks the optical sectioning capabilities of the much more popular f-SRM techniques working in the far-field regime, its importance is still very high as it does not require high laser beam power intensities (responsible for photobleaching and phototoxicity effects[64]), and can be used to investigate at nanoscale resolution any type of fluorescent sample ranging from biological species[65-67] (including autofluorescent ones), to nanostructured materials[68,69] in very simple illumination configurations relying on unsophisticated laser sources. Its resolution outputs depend on the size of the tip and 10-20 nm capabilities have been demonstrated even from its earliest variants[13,70]. An important advantage of TEF over f-SRM consists in the fact that it can be easily performed in tandem with other techniques based on tip-scanning, such as s-SNOM, PiFM, TERS, etc., resulting in spatially and temporally co-localized information. The combination of TEF with such complementary techniques can significantly boost the in-depth understanding of fluorescent samples. For example, in a correlative imaging approach, Meng and co-workers[71] theoretically proposed a nanoplasmonic strategy for precision in-situ measurements of TERS and TEF, which holds potential for becoming a popular approach for spectral analysis at nanoscale.

Although the field of TEN is gaining more momentum each day, most imaging & spectroscopy applications reported daily rely on commercial scanning probes that are not specifically developed for tip-enhanced nanoscopy, but for Atomic Force Microscopy (AFM)[11,72,73] or various other types of non-optical Scanning Probe Microscopy techniques. A consequence of this is that the optical properties of these tips are not provided by the manufacturers (being irrelevant for the applications they natively target). This issue is further accentuated by the fact that important properties such as absorption, scattering or reflectance are difficult to experimentally measure for objects at the scale of scanning probe tips. In this work, we propose a radically new approach for tip-enhanced nanoscopy, where techniques such as those mentioned above, harness the versatility, potential and performance of tips that are specifically designed and manufactured for applications in nanoscale tip-enhanced optical imaging. To this end, we propose a framework based on COMSOL simulations[74] and nano-manufacturing that represents a solution for developing specialized tips for TEN, with well-understood and tailorable optical properties. This follows up on recent related endeavors, which have demonstrated the very high-potential of TEN imaging with application customized probes[14,75]. We provide a proof-of-concept example where we design and fabricate two variants of silicon scanning probe tips featuring on their apex Au NPs, whose optical properties we approximated by numerical simulations. We argue that such probes can be useful in enabling nanoscopy applications that exploit in a well-understood manner the plasmon energy transfer occurring between the excited tip and the investigated samples, which can potentially enable higher signal-to-noise ratio either by direct mechanisms of background suppression or via indirect mechanisms based on the ultra-fast, and precise modulation of the tip's near-field. Furthermore, tunning the illumination wavelength to achieve different scattering profiles from the tip, in a well understood manner, can potentially be used in the modulation in near-field techniques whose contrast is intertwined with the tip's scattering. Additionally, precise knowledge on the tip's absorption and scattering properties holds great potential for optimizing existing nano-lithography methods based on tip-illumination, and for enabling novel tip-enhanced ablation/writing schemes. In the next, we describe our tip fabrication and modelling methodology, present our results, and provide a series of discussions on potential utility and applications.

## 2. Methods

### 2.1 Tip-fabrication methodology

Two custom-modified probes were synthesized, based on commercially available ones. We opted for silicon probes that do not present plasmon effects overlapping with those of the plasmonically active material that we employed (Au). Using such probes as foundation for this framework, results in a simple and reproducible fabrication process. In Fig. 1, SEM pictures of the two AFM tips, NP-S10 (Bruker) (Fig. 1(a)) and TESP-HAR (Bruker) (Fig. 1(b)), can be observed. These were custom modified (Fig. 1(c) and 1(d)), so that an Au NP is positioned on their apex, in two steps. The tips were coated by a 10 nm thick gold layer by sputter deposition. Second, the gold layer was removed by Focused Ion Beam (FIB)[76,77] sputtering on the whole tip, except from its apex, resulting in an Au nanoparticle (NP) being left there. In Fig. 1(c) and 1(d), one can observe the last step of the tip fabrication.

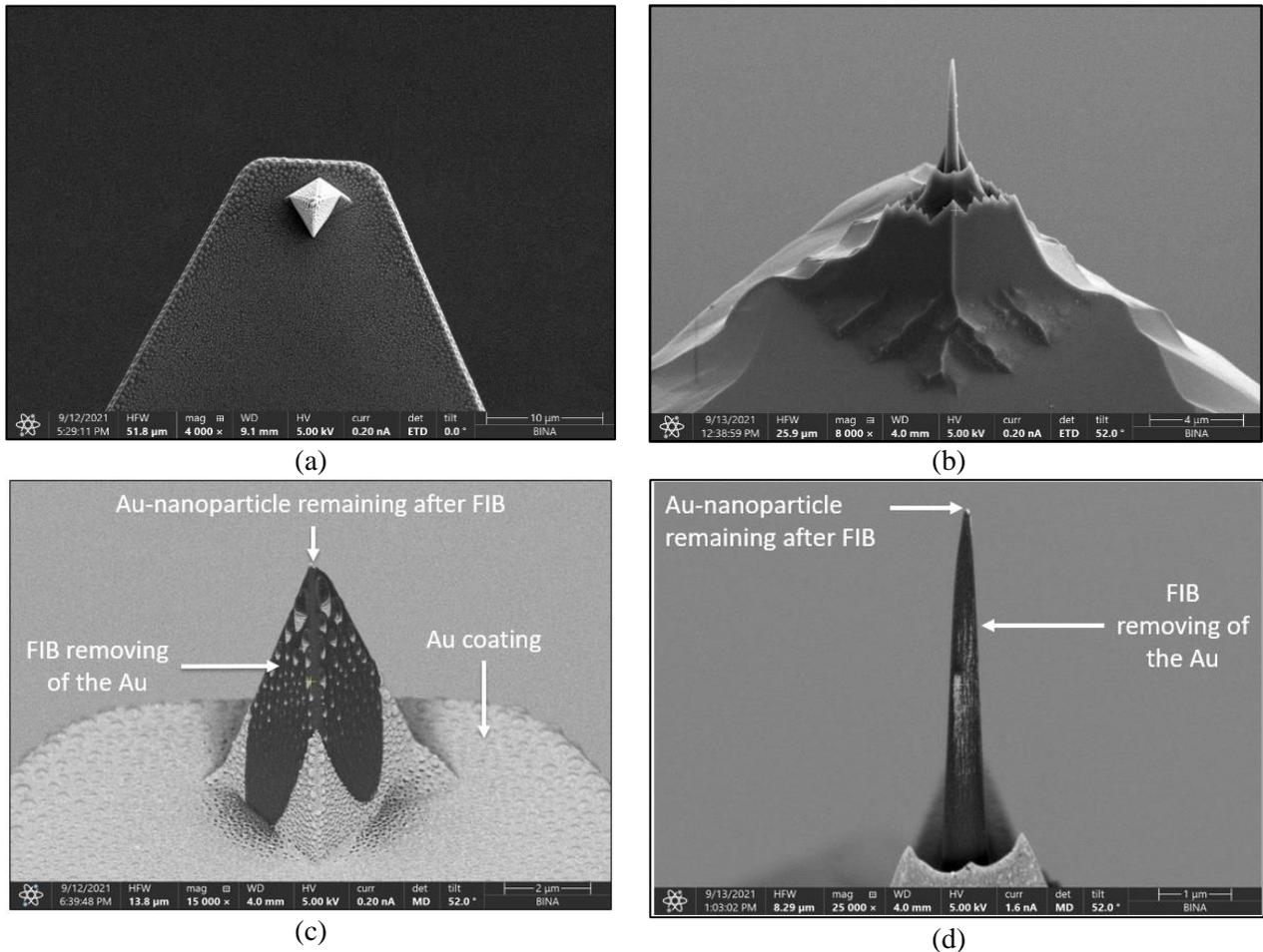

**Fig. 1.** SEM micrographs of the commercial AFM probes before and after custom-processing. (a) NP-S10 probe (bottom view); (b) TESP-HAR probe (side view); (c) NP-S10 probe after Au thin film deposition and subsequent FIB processing. FIB removal of the gold thin film can be observed; (d) TESP-HAR probe after Au thin film deposition and subsequent FIB processing.

## 2.2 COMSOL simulations methodology

2.2.1 Half-sphere and cylinder models

The physical parameters of interest in this experiment are the probe's plasmon band and its size. While experimental measurement on the scattering and absorption spectrum of the probe could potentially be used to assess the plasmon band of the probe, such results are very difficult to obtain on nanosized objects. Numerical analysis does not suffer of this limitation. Moreover, as recently shown also in other works on TEN imaging, simulation methods allow testing of different tip geometries and sizes, so as to identify their influence on the imaging outputs[78]. As discussed further in this section, to fabricate TEN probes with known optical properties, we adopted a fabrication approach whose final output is the deposition of an Au NP of custom-size on the apex of a commercial AFM probe. Two commercial AFM probes with distinct geometries were considered, NP-S10 (Bruker) and TESP-HAR (Bruker). By simulating the optical properties of this NP, we argue that the optical properties of the tip can be (partially) approximated.

In Fig. 2(a), a scanning electron microscope (SEM) picture presenting the apex of one of the custom-modified probes is presented. At the extremity of the tip, the Au NP can be clearly noticed. According to the fabrication process, discussed earlier, the NP deposited on the probe's apex shares an ellipsoidal shape. However, for the sake of simplifying the modelling work, its shape was considered spherical, with a radius of R. The optical properties of this particle are those responsible for the potential Plasmon Resonance Energy Transfer (PRET) effects occurring from tip to sample. Given that we have used a probe substrate (Si) that is plasmonically inactive in the spectral region matching the plasmon band of the Au NP deposited on the tip apex, we restrict to considering the properties of this Au particle in the COMSOL models used for simulating the tip's optical properties. In this regard, two geometrical models are proposed and used to match with the NP's shape: a half-sphere (Fig. 2(b)) and a thin cylinder (Fig. 2(c)).

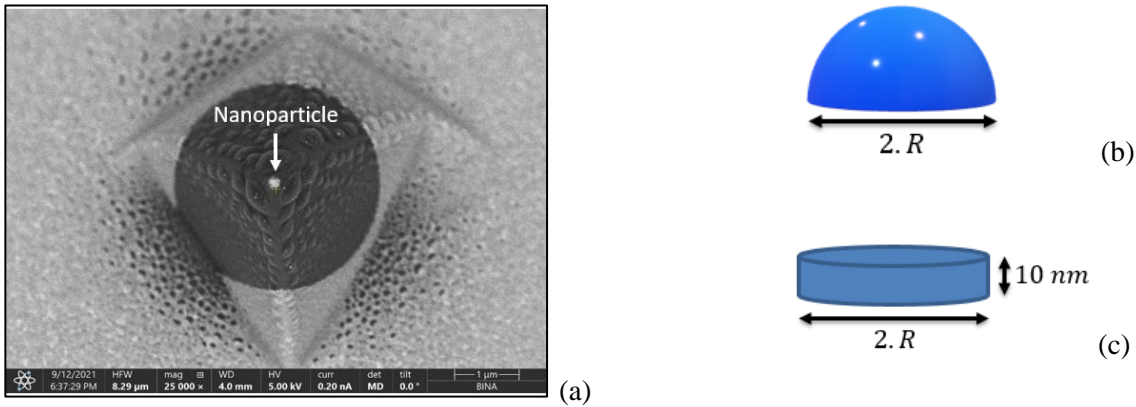

**Fig. 2.** (a) SEM image of the custom-modified probe (NP-S10): AFM Silicon tip with an Au NP at its top. One can assume this NP circular of radius R; (b) Diagram of half-sphere; (c) Thin-cylinder by which the NP can be approximated.

According to measurements performed after the fabrication procedure, the *R* radius of the NP deposited on the probe's apex lies in the following intervals:

- For the NP-S10 based tip: 129.5 nm < 2R < 153.8 nm
- For the TESP-HAR based tip: 70.15 nm < 2R < 75.54 nm.

These results were used in simulations, to approximate the optical properties of the NP positioned on the tip apex. The initial radii of curvature of the two considered probes were, 20 nm, and 10 nm for the NP-S10 and TESP-HAR probes, respectively.

2.2.2 Models implementation in COMSOL

Given the high computational time required for the simulation of the optical properties of a 3D gold sphere, only a quarter of the structure was designed, whereas in the calculations of the results the software extrapolates the achieved results to the entire structure. This methodology was inspired from a COMSOL example on Mie Scattering addressing structures of similar kind. In Fig. 3, one can discern three different layers: the most external one is a Perfectly Matched Layer (PML), the middle-inner layer represents an air layer, and the most internal layer represents the gold NP. The PML is used to simulate open boundaries[79] by absorbing the outgoing electromagnetic waves and for canceling the reflection effect. The illumination, simulated as electromagnetic waves, is considered along the x-axis. For numerical convenience, only half of the particle geometry structure was designed, whereas the influence of the remaining part could be obtained by simple calculation thanks to the particle symmetry. The measurement of the scattering and of the absorption were performed by integration over the entire external interface of the air layer. In Fig. 3(a) we present COMSOL geometrical structure for the half-sphere model, and in Fig. 3(b) for cylinder model.

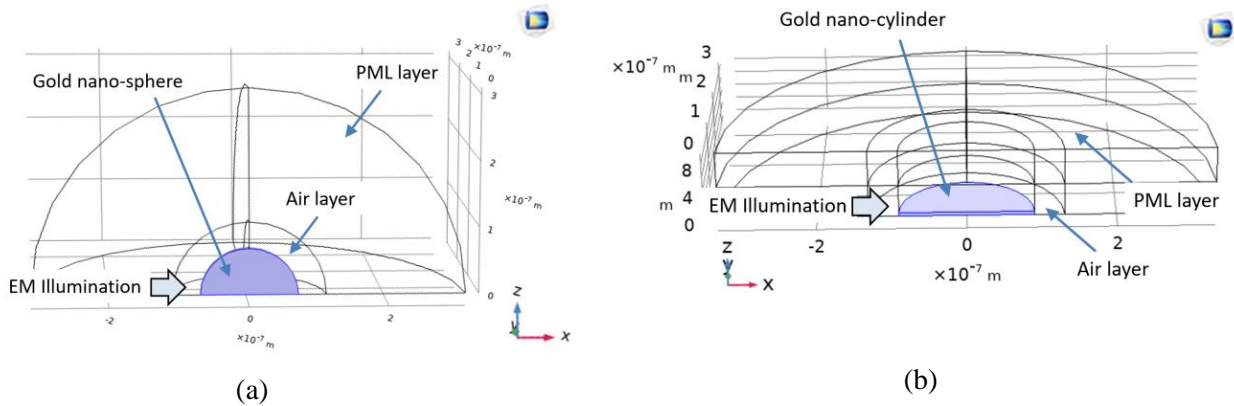

**Fig. 3.** COMSOL geometry models. (a) Half-sphere model; (b) Thin-cylinder model. The geometry is divided in three layers: PML, air and Au NP. The illumination of the sphere in considered along the x-axis.

## 2.3 Resonance characterization methodology

To measure the effective plasmon resonance of the custom-modified probes, spectrometer measurements were designed. A challenging experimental setup was used for this purpose. The method consists in measuring the spectrum of the scattered light of the tip using a spectrometer connected to fiber optics. The scattering of the tip is obtained using dark field microscopy through a Carl Zeiss Axiotron optical microscope, and a ThorLabs Compact Spectrometer 350 - 700 nm was connected to the system, and used for the resonance measurements. Fig. 4 presents a series of cantilever and tip pictures under bright (Fig. 4(a)) and dark (Fig. 4(b) and 4(c)) field microscopy. Fig. 4(d) shows the experimental system configuration.

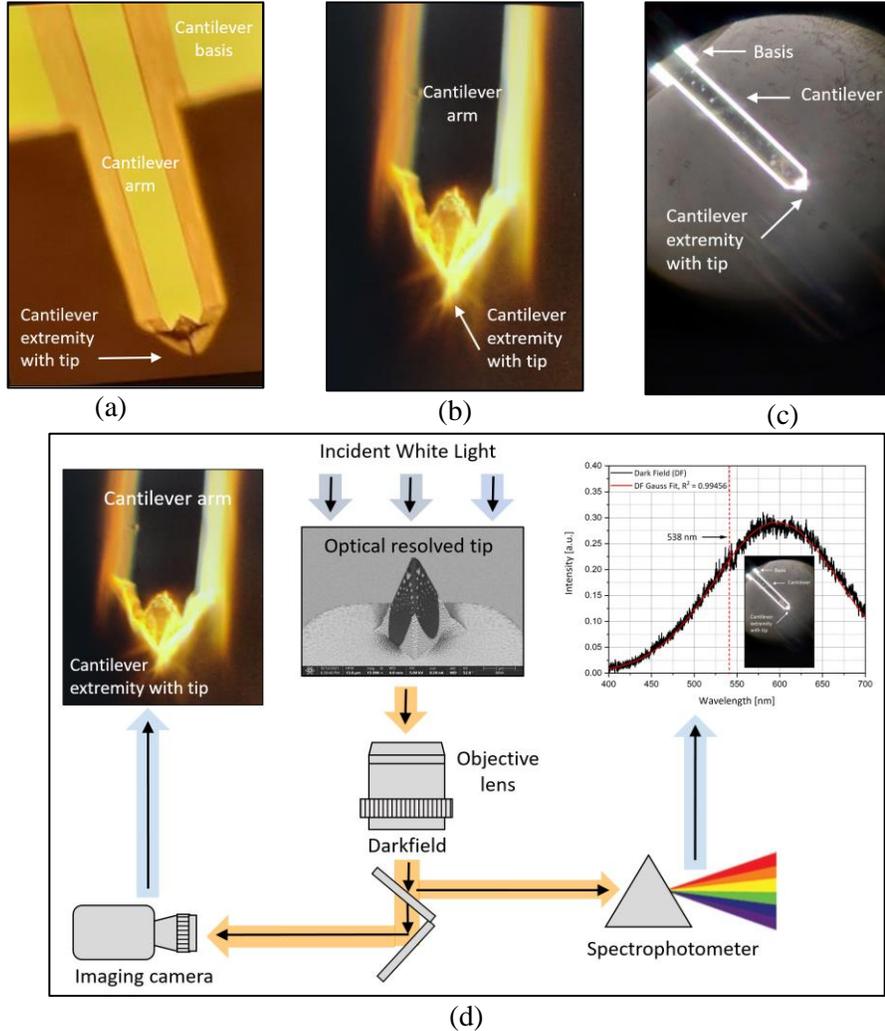

**Fig. 4.** Cantilever microscope magnified pictures. (a) Cantilever extremity under bright field; (b) Cantilever extremity under dark field; (c) Full cantilever under dark field; (d) Experimental system configuration.

## 3. Results

### 3.1 Au apex shapes and measured dimensions

As discussed in the Introduction section, the purpose of this experiment is to design a methodology combining nano-fabrication and numerical simulations that can result in probes sharing known optical properties, which can be used to augment TEN imaging outputs. For this purpose, commercial grade probes were custom-modified to equip their apex with Au NPs, whose optical properties can be easily approximated by numerical simulations. In Fig. 5 one can observe the dimensions of the Au NPs resulted for the NP-S10 (Fig. 5(a)), and TESP-HAR (Fig. 5(b)) probes, respectively, following the FIB processing procedure described under the Methods section. Regarding the NP-S10 probe, the smallest NP obtained, according to the FIB resolution limitation, is of ellipsoidal shape, with 129.5 nm and 153.8 nm, on the two axes. The eccentricity $e$ of this ellipsoid is equal to 0.54. In the simulation part, we approximated a spherical shape for this NP, and a range of radii between the semi-minor and semi-major axis. Regarding the TESP-HAR probe, the smallest Au NP obtained was also of ellipsoidal shape, this time with 70.15 nm and 75.54 nm on the two axes. The eccentricity $e$ of this ellipsoid is equal to 0.27. Fig. 5(c) and 5(d) present top views of the different existing diameters from the bottom of the tip to its top.

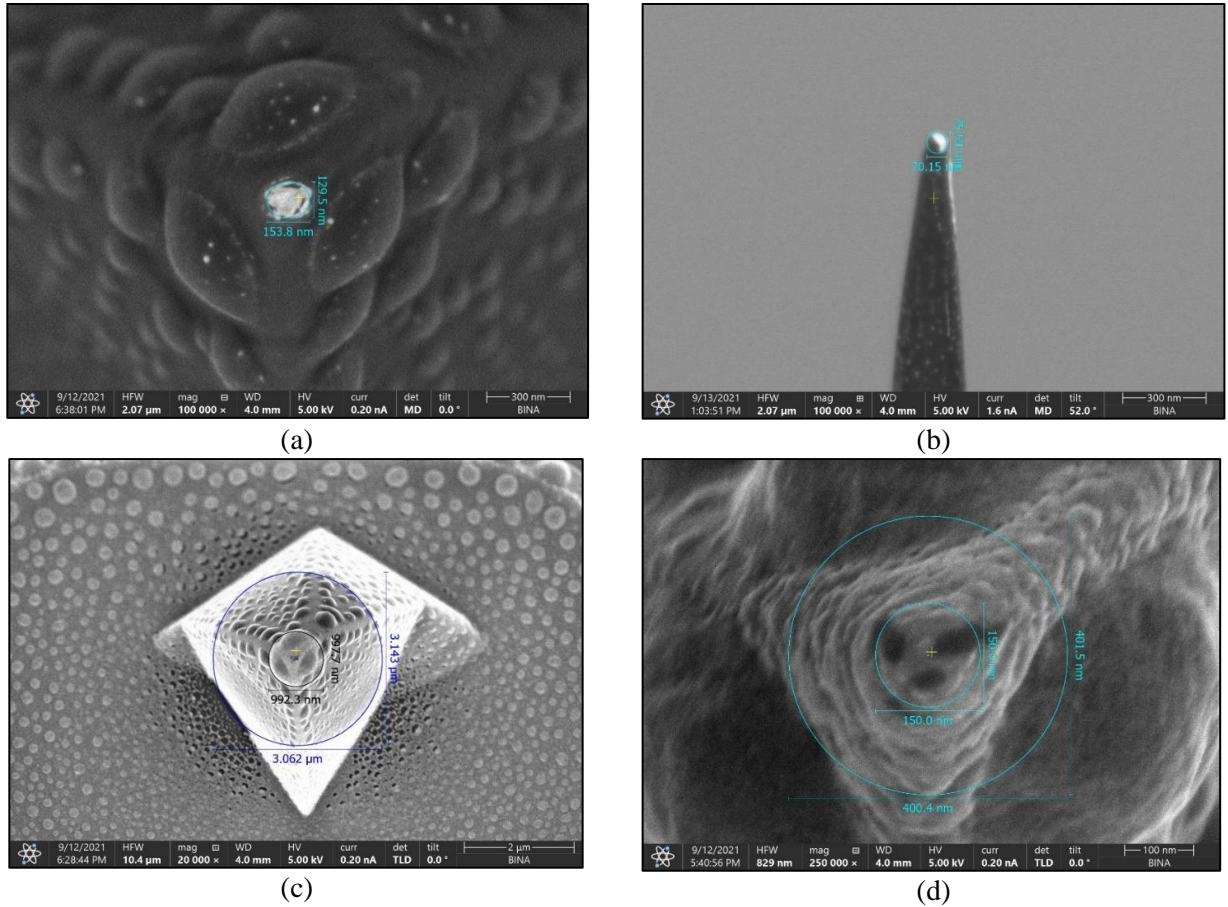

**Fig. 5.** SEM pictures of the custom-modified tips. Au NPs can be observed on the apex of the tip. (a) NP-S10 probe; (b) TESP-HAR probe; (c) Top view of the diameters starting from the basis of the tip (μm) and up towards upper stages of the tip (nm); (d) Zoom-in of the top view. Dimensions of Au NPs are presented in (a) and (b).

## 3.2 Numerical results

3.2.1 COMSOL simulation results for the half-sphere model

In Fig. 6, we present the simulations of the scattered cross section (SCS) of the field (Fig. 6(a) and 6(c)), and the absorbed cross section (ACS) of the field (Fig. 6(b) and 6(d)), as a function of the wavelength, for the half-sphere model. Different radii $R$ (Fig. 2) were simulated, considering dimensions close to the radius $R$ of the fabricated tips. Since in reality the nanoparticle exhibits an ellipsoidal shape, different radii were tested in the range of the semi-minor axis to the semi-major axis. We assume that the properties of the real tips are close in value to the results numerically simulated for the considered radii range. In Table 1, the value of the SCS and ACS maxima are summarized as a function of the radius.

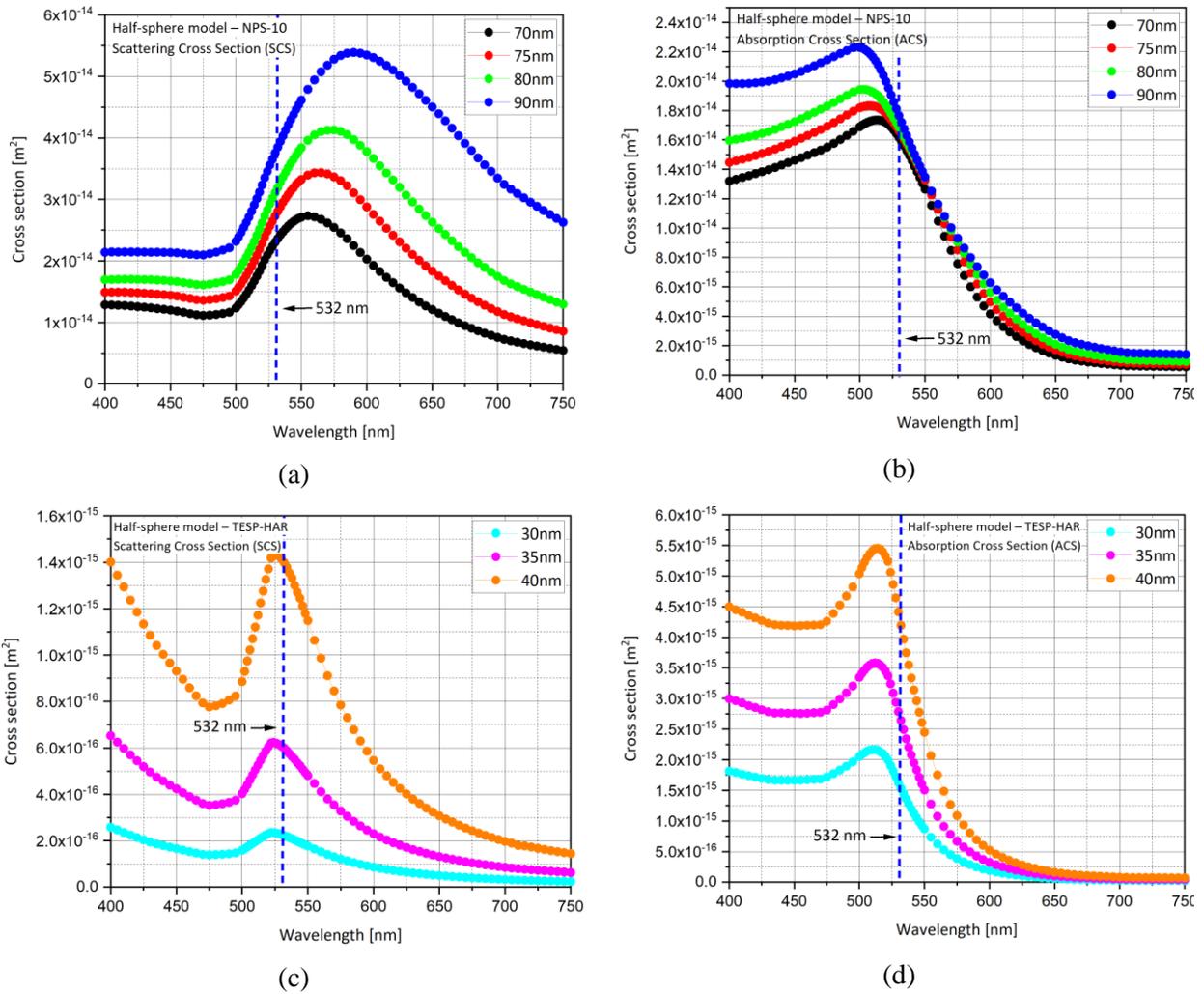

**Fig. 6.** COMSOL results on the SCS and ACS using the half-sphere model. For NPs of radius matching the dimensions of the NP deposited on the apex of the NP-S10 probe: (a) Graph of the scattering; (b) Graph of the absorption. For NPs of radius matching the dimensions of the NP deposited on the apex of the TESP-HAR based tip: (c) Graph of the scattering; (d) Graph of the absorption.

Table 1. Wavelength of the maximum scattering and absorbance for radii of different sizes, calculated by approximating the tip shape with the half-sphere model.

| Radius | 30 nm | 35 nm | 40 nm | 70 nm | 75 nm | 80 nm | 90 nm |
|---|---|---|---|---|---|---|---|
| Scattering | 522 nm | 524 nm | 526 nm | 555 nm | 565 nm | 575 nm | 590 nm |
| Absorbance | 512 nm | 512 nm | 514 nm | 514 nm | 508 nm | 502 nm | 500 nm |

For radius $R$, corresponding to the ellipsoidal nanoparticle of the NPS-10, one can observe the following: First, for all the considered radii, a local maximum of the scattering is observed between 555 nm to 590 nm (see also Table 1). These values are close to 532 nm and 561 nm, two wavelengths that are commonly used for fluorescence microscopy[80,81], an aspect that can potentially be used in TEF applications building on plasmon resonance energy transfer. Second, the wavelength corresponding to the resonance of the scattering increases with the NP radius. Concerning the ACS, for all the different radii, the maxima are observed between 500 nm to 514 nm (see Table 1). In Fig. 6(c) and Fig. 6(d) we present the SCS and ACS results, respectively, for the considered radii range corresponding to the ellipsoidal nanoparticle of the TESP-HAR probe. The SCS maxima were obtained in the range of 522 nm to 526 nm, while the ACS maxima were obtained in the range of 512 nm to 514 nm.

According to the Mie theory, the two bands have to match. The corresponding scattering and absorption cross-sections of a sphere of radius $a$, calculated via the Pointing vector, from the full electro-magnetic (EM) field, associated with an oscillating dipole, are defined with a set of equations[82]. While the scattering cross-section expression is given by equation (1):

$$\sigma_{sca} = \frac{k^4}{6\pi}|\alpha|^2 = \frac{8\pi}{3}k^4 a^6 \left|\frac{\varepsilon(\omega)-\varepsilon_d}{\varepsilon(\omega)+2\varepsilon_d}\right|^2 \tag{1}$$

The absorption cross-section is given by equation (2):

$$\sigma_{abs} = kIm(\alpha) = 4\pi k a^3 Im\left[\frac{\varepsilon(\omega)-\varepsilon_d}{\varepsilon(\omega)+2\varepsilon_d}\right] \tag{2}$$

The total cross-section is given by equation (3):

$$\sigma_{ext} = \sigma_{sca} + \sigma_{abs} \tag{3}$$

Where:

- $\alpha = 4\pi a^3 \frac{\varepsilon(\omega)-\varepsilon_d}{\varepsilon(\omega)+2\varepsilon_d}$
- $\omega =$ frequency of the planar wave illumination
- $\varepsilon =$ Permittivity of the gold nanoparticle.
- $\varepsilon_d =$ Permittivity of the dielectric (air).

Then the resonance will be obtained through equation (4):

$$I_{ext}(\omega) = \frac{I_0(\omega)}{S}\sigma_{ext}(\omega) \quad (4)$$

In case of a large NP, the corresponding scattering and absorption cross-sections will become equations (5) and (6) respectively:

$$\sigma_{sca} = \frac{2\pi}{k^2}\sum_{n=1}^{\infty}(2n+1)(|a_n|^2 + |b_n|^2) \quad (5)$$

and

$$\sigma_{abs} = \frac{2\pi}{k^2}\sum_{n=1}^{\infty}(2n+1)Re(a_n + b_n) \quad (6)$$

so we get:

$$\sigma_{abs} = \sigma_{ext} - \sigma_{sca} \quad (7)$$

In such a case of large particles, the corresponding cross-sections are still calculated via the Mie theory, while the electrostatic approximation breaks down, and the retardation effects are to be considered. In case of PRET[83], a process that can influence tip-sample interactions accounting for TEN contrast (e.g. in tip-enhanced fluorescence), the energy absorbed by the tip, via plasmon excitation, is transferred to the sample. In this case, the ACS is more relevant than the SCS. In the case of s-SNOM, the scattered field comprises a series of terms[84], namely, the incident field (i) scattered by the tip, (ii) scattered by the sample, (iii) scattered by the sample and then by the tip, (iv) scattered by the tip and then by the sample, and (v) scattered by the tip, then the sample, and finally the tip again. The sample properties dictate the weight of each of these terms in the recorded signals, but for an s-SNOM configuration based on detection at higher harmonics of the tip's oscillation frequency, the incident field scattered by the sample should be discarded[84]. Thus, considering that the scattering by the tip is mainly responsible for the s-SNOM attainable contrast, in potential s-SNOM applications with optically resolved probes the SCS may be more relevant than the ACS. In nanolithography[85] and ablation[86] approaches based on tip-enhanced illumination both ACS and SCS can be equally important for optimizing the final outputs, depending on the specifics of the application at hand.

3.2.2 COMSOL simulation results for the cylinder model

In Fig. 7, we present the simulated SCS and ACS as a function of the wavelength, for the cylinder model. For *R* radius, corresponding to the ellipsoidal nanoparticle of the NPS-10, for the SCS (Fig. 7(a)) and for the ACS (Fig. 7(b)), the following phenomena can be observed: First, for all the different radii, a local maximum of the scattering phenomenon is observed between 522 nm to 530 nm. However, contrary to the half-sphere model results, one can notice that the wavelength corresponding to the resonance of the scattering seems here to decrease with the radius. Concerning the ACS, for all the different radii, the maxima are observed between 500 nm to 504 nm. The SCS and ACS results for the NP deposited on the apex of the TESP-HAR are presented in Fig. 7(c), and Fig. 7(d), respectively. In this case, the SCS maxima are localized at between 530 nm to 534 nm, whereas the ACS maxima are localized between 508 nm to 518 nm. The value of the SCS and ACS maxima are summarized in Table 2 as a function of the NP radius.

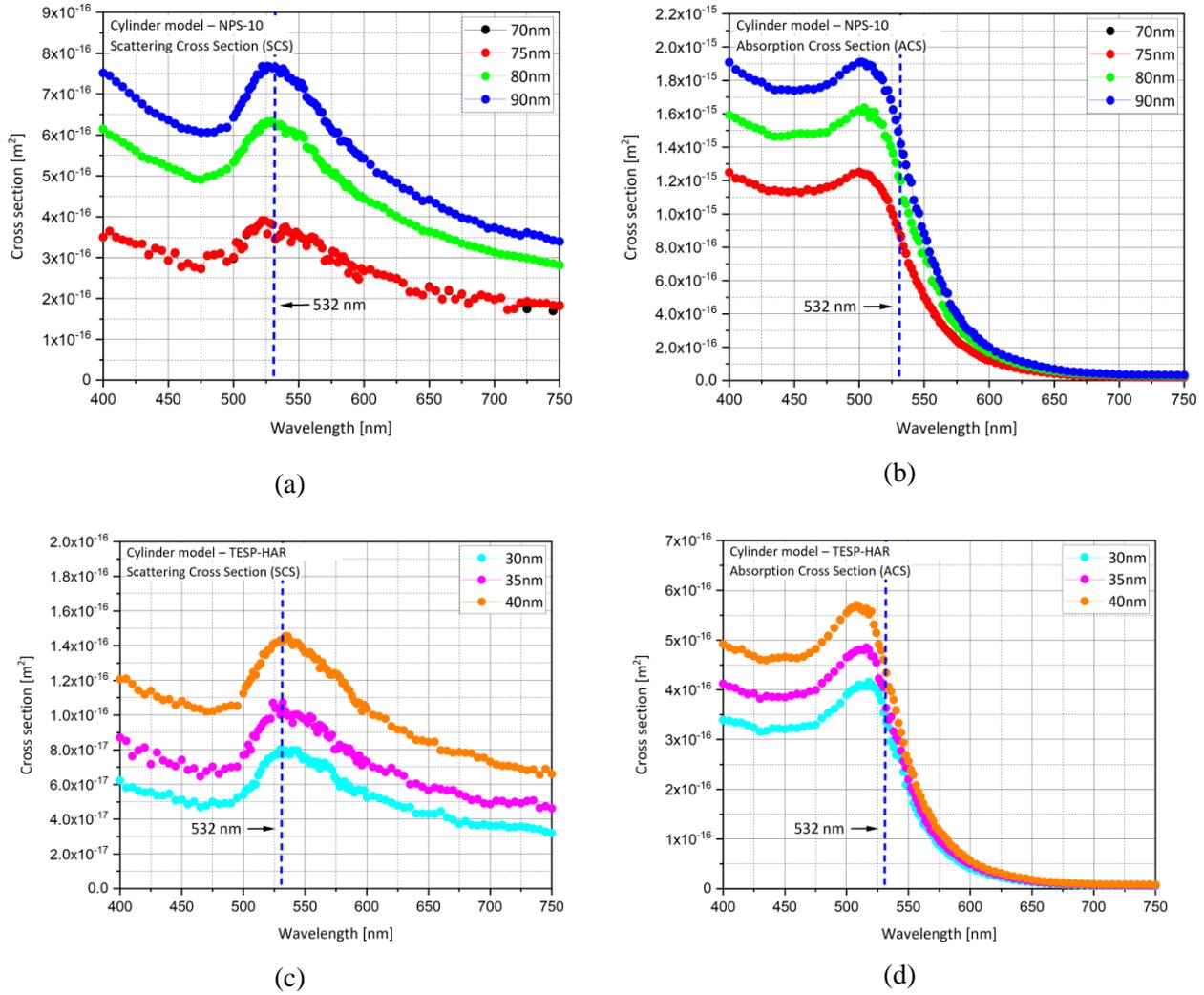

**Fig. 7.** COMSOL results on the SCS and ACS using the cylinder model. For NPs of radius matching the dimensions of the NP deposited on the apex of the NP-S10 probe: (a) Graph of the scattering; (b) Graph of the absorption. In this case, 70 nm and 75 nm curves are fully matched. For NPs of radius matching the dimensions of the NP deposited on the apex of the NP-S10 TESP-HAR probe: (c) Graph of the scattering; (d) Graph of the absorption.

**Table 2.** Wavelength of the maximum scattering and absorbance for radii of different sizes, calculated by approximating the tip shape with the cylinder model.

| Radius | 30 nm | 35 nm | 40 nm | 70 nm | 75 nm | 80 nm | 90 nm |
|---|---|---|---|---|---|---|---|
| Scattering | 530 nm | 532 nm | 534 nm | 522 nm | 522 nm | 530 nm | 524 nm |
| Absorbance | 518 nm | 516 nm | 508 nm | 500 nm | 500 nm | 504 nm | 502 nm |

## 3.3 Characterization results

### 3.3.1 Bright field vs. dark field

To enable clear identification of the plasmon response, and after checking the tip in bright field mode (Fig. 4(a)), it is now necessary to illuminate the cantilever and the tip in dark field mode (Fig. 4(b) and 4(c)). Several challenges rose up when trying to illuminate the tip. First, even if the illumination is off, the background illumination resulted from the scattering due to the edges of the Silicon made cantilever still has influence on the spectrum. Then, while trying to isolate the resonance peak from the extremity of the gold nano-particle, one can assume the additional spectral responses are obtained from the rest of the probe, also covered by metal layer. This is why it is not straight forward to measure such low photo-activated answers, and some gaps may appear between the simulations, where one can isolate an illuminated surface with high accuracy, and the measured spectrum obtained through the whole system.

### 3.3.2 Spectrometer curves

While numerical results (a.k.a. COMSOL simulations) presented a forecast of the tips' expected behavior, experimental results became necessary to confirm and analyze the scattering (SCS), and the absorption (ACS) phenomena. As per previous works, based on Mie theory, it appeared that for Si highly-doped NPs (a.k.a. acting as metal), the resonance frequency increases with the size of the NP[87]. Similar conclusions were obtained with silver NPs sharing different sizes[88]. Fig. 8 presents a series of measured spectra, while illuminating the tips in bright and dark field modes. At this stage, it appeared that the most promising type of tip, towards future optical resolved scanning, was the TESP-HAR model, since the obtained results were more solid. Fig. 8(a) presents the measured spectrum in bright field mode. A resonance peak can be observed at 538 nm. Fig. 8(b) presents the measured spectrum in dark field mode. Although more dim, the resonance peak at 538 nm can still be observed. It should be noted that the measured intensity in dark field is almost one third when compared to bright field mode's intensity, for same integration time of 900 nm.

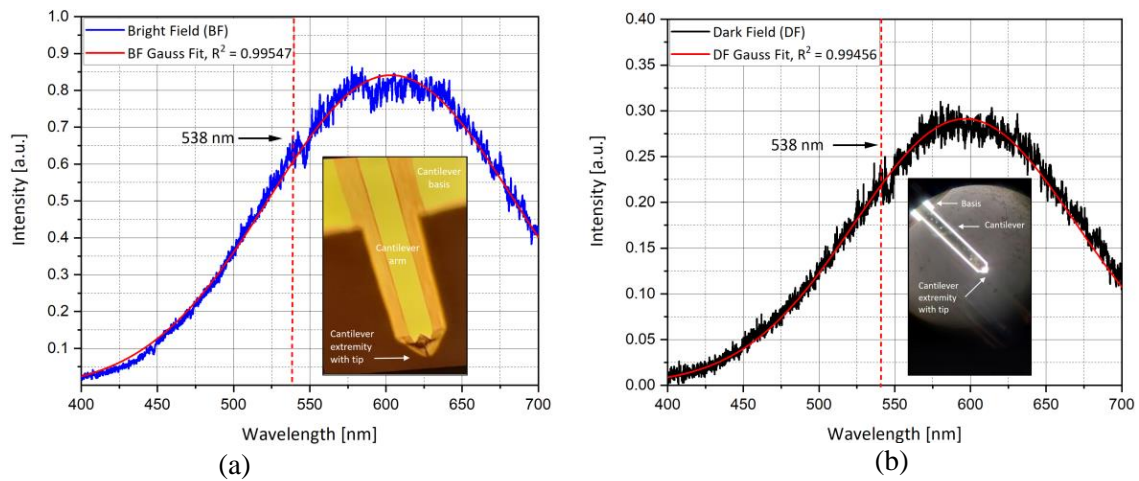

(a)　　　　　　　　　　　　　　　　　(b)

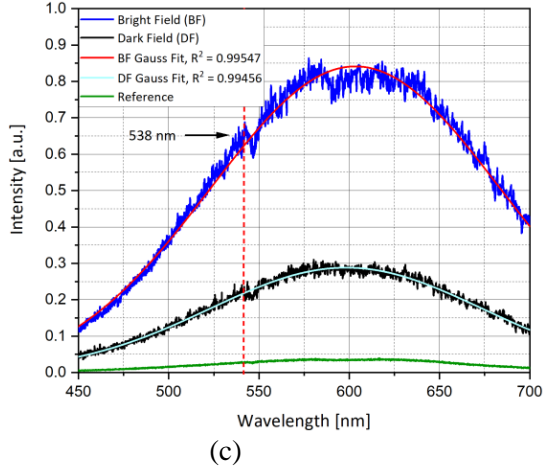

(c)

**Fig. 8.** Spectrometer measurements of the resonance wavelength. (a) Bright field mode. Magnification M = ×150 and integration time t = 900 ms; (b) Dark field mode. Magnification M = ×50 and integration time t = 900 ms; (c) Comparison of bright field, dark field and reference spectra.

## 4. Discussion

At present, although TEN imaging has enabled important discoveries, the outputs of these techniques are largely biased by the use of non-specialized probes. Here we propose the idea of using optically resolved probes, which can potentially significantly augment nanoscale imaging with TEN, by enabling novel routes for higher SNR, and consequently, higher resolution and faster acquisition rates. We showed that by coating a commercial AFM probe with a thin layer of gold, and then processing it by FIB, an Au NP can be placed on its apex. By experimenting with two, very different probes, we demonstrated that this methodology can be implemented irrespective of the tip's geometry. Other materials besides Au, such as Pt, Cr, Ag, etc. could also be used, depending on the tip properties required by specific applications. Alternative approaches, not demonstrated here, could consist in attaching NPs of known size and shape to tipless cantilevers, but according to our past experiences on the topic, such approaches can be significantly more challenging. Furthermore, such approaches would result in probes with peculiar tip-shanks, whereas current body of understanding on near-field data[78] incorporates the specific geometries of traditional AFM tips, which the methodology proposed here, better maintains. Further on, we showed that by COMSOL numerical simulations we can approximate the optical properties of the NP resulted on the probe apex after FIB processing. Experimental assays based on bright-field and dark-field microscopy and spectroscopy, can be eventually implemented to validate the numerical results. We consider that the proposed framework could result in the advent of novel TEN imaging approaches where investigations are done with specialized tips whose known optical properties can be exploited for better and faster investigations. To this end, in the next paragraph we discuss two potential applications.

TEF implementations that rely on the coupling between the absorption band of fluorophore and localized surface plasmon resonance of the tip-substrate system [13,89,90], hold important potential to resolve many of the typical drawbacks of more simplistic TEF imaging approaches, where the properties of the probe are not thoroughly taken into account [65,66,91], such as: low signal to noise ratio, and intrinsic high-acquisition time. In traditional tip-enhanced fluorescence microscopy all fluorophores under the influence of the laser beam tuned to the absorption band of the fluorophore beam emit light. The light emitted by a single fluorescent particle (or by a cluster of reduced size) is modulated by vertically oscillating a metallic tip on top of it and extracted with a lock-in amplifier (Fig. 9(a)). Intense background accounts for severe drawbacks such low SNR, and consequently low acquisition time, which hinders the popularity of such imaging approaches, despite their valuable potential for ultra-high resolution. In an alternative approach that can potentially overcome these drawbacks, the fluorophores are excited via the plasmon resonance energy transfer[83,92] from a tip that is illuminated with a wavelength promoting surface plasmon resonances in it. This can lead to a significant suppression of background signals, as the fluorophores close to the excited one do not benefit of the PRET effect (Fig. 9(a)). Implementing this imaging modality requires that the plasmon band of the tip and the absorption band of the probed fluorescent molecules overlap[83], Fig. 9(b). However the plasmon band of commercial AFM probes is not explicitly provided by manufactures, nor can be easily measured, thus approaches as above described can currently be implemented based on trial and error approaches, which is impractical. Our results demonstrate the valuable synergy between a numerical simulation framework and a nano-fabrication methodology, which can be used to obtain tips with known plasmon band, which are highly suitable for the above-described application. Specifically, the developed tips have a well-defined absorption band, with ACS maxima ranging between 500 nm and 518 nm, depending on the size of the NP resulted on the tip apex after FIB processing (Tables 1 and 2), as was assessed by numerical simulations. Illuminating in the simulated absorption band, leads to local surface plasmon resonances in the tip apex, which can be transferred to the sample via a plasmon resonance energy transfer effect, supposing the probed molecules have an absorption band matching the plasmon band of the tip.

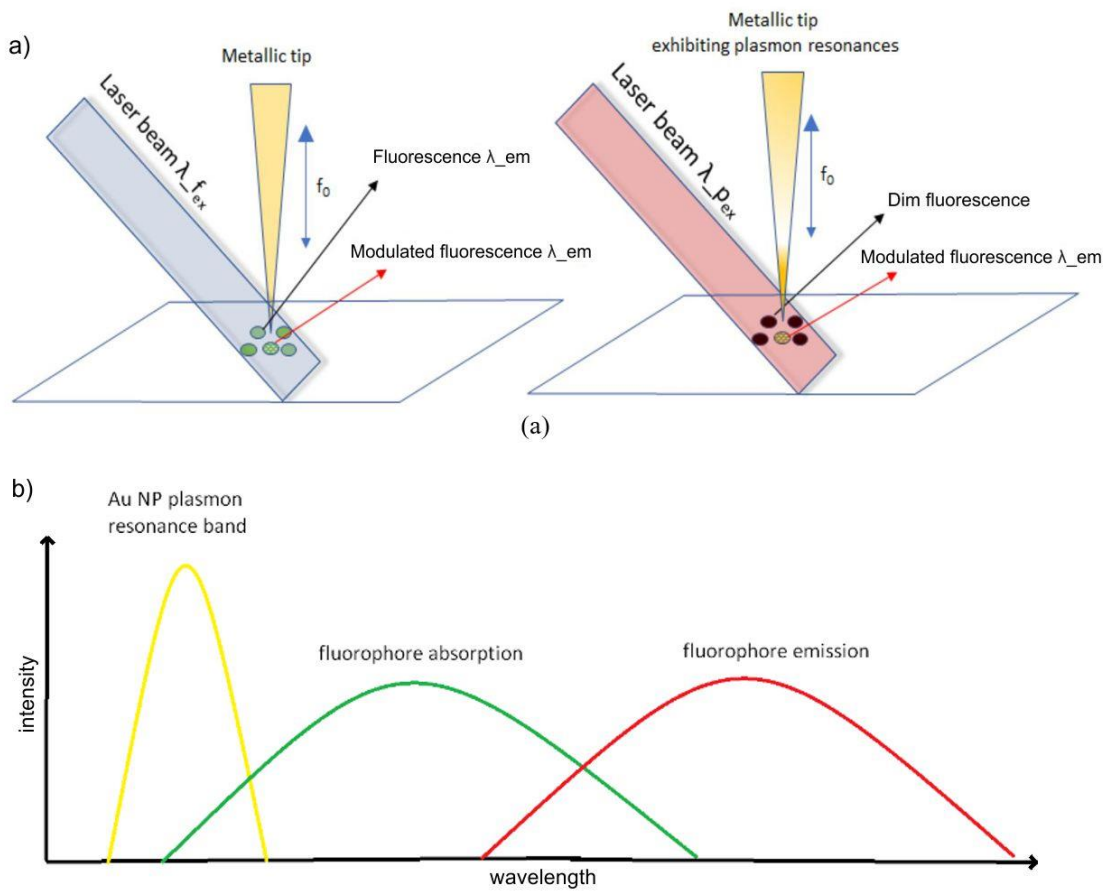

**Fig. 9.** (a) Conventional tip-enhanced fluorescence (left) vs. tip-enhanced fluorescence based on plasmon resonance energy transfer (right). In conventional TEF, the optical properties of the tip are not considered, and the excitation ($\lambda\_f_{ex}$) is tuned to the maximum of the absorption band of the fluorescent molecule of interest. With PRET based TEF, the laser beam used for excitation ($\lambda\_p_{ex}$) matches the plasmon band of the tip, and fluorescence is obtained by dual effect, direct illumination and plasmon resonance energy transfer from tip (donor) to sample (acceptor), thus reducing the background and yielding higher SNR; (b) Condition for the PRET effect to take place. The plasmon resonance band overlaps with the fluorophore absorption band. When the plasmon band is positioned on the left tail of the absorption spectrum, the excitation light has a low contribution on the emission of illuminated fluorophores not benefitting of the PRET effect. These latter would provide stronger emission due to the dual PRET + direct excitation effect.

The following question naturally arises? Would investigations on distinct fluorophores, with different absorption bands, require the use of different tips, each with a plasmon band matching the absorption band of the fluorophore in question? We speculate that this can potentially be avoided by implementing clever strategies where the absorption/plasmon band of the tip can be tuned. Such a potential strategy can eventually rely on previous work showing that sub-micron particle-based structures can be employed to reconfigure photonic devices by external photonic radiation. One of the device configurations that was previously demonstrated involved metallic nanoparticles or nanorods incorporated into a semiconducting structured matrix that exhibits strong plasmonic resonance which depends on the power of the external photonic radiation that is being absorbed by the matrix[93]. Taking inspiration from this previous works, future specialized TEN probes can be designed to feature optical tunability obtained by adding an external control beam that when absorbed in a silicon part of the tip nanostructure will produce free electrons leading to tuning of the plasmonic resonance of the nano structure at the end of the tip and by that improving the energy transfer process by properly matching the plasmonic resonance to the plasmonic excitation wavelength. Given that such tips will possess a tunable plasmon band, the same tip will be capable to image various fluorescent molecules. Besides its use for fluorescence imaging, this strategy could also enable sub-nanometer imaging of non-fluorescent materials.

On a different train of thought, we refer to probably the most notorious TEN techniques, s-SNOM. The working principles of s-SNOM rely on a sharp tip that is scanned across the sample's surface while it is being excited with a focused laser beam. The attainable resolution in s-SNOM depends on the size and geometry of the probe, with many important applications having been reported for tip sizes ranging between 5 and 150 nm[73,79,94,95]. The tip converts the incident radiation into a highly localized and enhanced near field at the tip apex, which modifies both the amplitude and the phase of the scattered light via the near-field interaction with the sample underneath[96]. This process depends on the local dielectric properties of the tip and sample. The default way of extracting s-SNOM signals is by lock-in detection of the optical signals corresponding to the tip-sample interaction, which are modulated by the tapping movement of the tip[96]. We speculate that this type of modulation, achieved by mechanical means, can be replaced, or augmented with an alternative/additional modulation scheme, by means of a pump beam tuned to the plasmon band of the tip, which can intermittently plasmonically activate/deactivate the tip, at ultra-fast rates, and depending on the properties of the sample, it can modify the interaction of the two mirror dipoles (in probe and sample) accounting for s-SNOM contrast[97,98]. This pump beam, would complement the probe beam, which in many s-SNOM applications is tuned to the absorption band of the sample to promote phase contrast[36,72,96,99]. An additional layer of (faster) modulation can eventually provide higher SNR. Furthermore, this alternative modulation scheme based on on-ff switching of the pump beam, could even entirely replace the traditional modulation schemes based on the tapping movement of the probe, which could enable s-SNOM imaging in contact mode, which would be faster than tapping-mode s-SNOM (which is the current default way of performing s-SNOM investigations), and more suited for some applications, depending on the sample type and scanning environment. Besides considering the tip's absorption as the key for signal modulation, referring to its scattering can also represent a route. For example, by intermittently using two, close wavelengths, accounting for significantly different scattering from the tip, we speculate that the s-SNOM signals can be modulated as well. Similar modulation approaches can eventually be implemented for other

TEN modalities, such as TERS, PiFM, etc. Overall, we consider that having at hand probes with precisely known optical properties can open an entire new palette of data acquisition strategies for TEN techniques that can enable better outputs in terms of signal quality, sensitivity, robustness and speed.

Besides imaging, the interaction between focused light and AFM probes, was also demonstrated highly-useful to date for nano-lithography and ablation purposes[85,86,100]. We argue that in such applications the use of optically resolved probes can also be important in optimizing the outputs, as the interactions between tip and sample are likely to be entirely different for situations when the illumination wavelength matches or not the absorption or scattering of the tip. Furthermore, these properties can potentially also be correlated with the heating profiles of the tip, which were previously shown to have an important impact over the final outputs of tip-enhanced nano-ablation[101].

Last we feel important to turn our attention to the issue of attainable resolution. While the results presented here do not align, in terms of tip size, to the state-of-the art on AFM probe sharpness, the motivation behind this proof-of-concept experiment was not aimed to address this particular problem. We speculate that adjusting our proposed methodology to incorporate the use of probes with different geometries, in combination with different processing methods, can eventually result in sharper probes. However, we find noteworthy to recall recent results[73], which showed that tip-sharpness is not always preferred. Thus, we consider that having a precise understanding of the tip's optical properties, can, at times, be the better option to consider.

## 5. Conclusions

Tip-enhanced nanoscopies have enabled to date a wide palette of discoveries, and account for new scientific breakthroughs each day. However, they rely on non-specialized probes, usually designed and commercialized for various non-optical Scanning Probe Microscopy variants, which significantly biases their potential. Here, we propose the concept of TEN probes with known optical properties. To this end, we custom-modify two commercial AFM probes to equip their apex with an NP, whose absorption, and scattering properties we approximate by numerical simulations in COMSOL based on Mie scattering[102,103]. We argue that the so developed probes, with resolved optical properties, can enable various novel data acquisition schemes for faster and more sensitive TEN imaging, and may also provide useful for nano-lithography A series of perspectives in this regard are discussed. Overall, we consider the reported framework to be highly flexible, allowing the design and fabrication of a wide range of tips with custom-tailorable optical properties exhibiting various geometries and materials compositions.

**Disclosure**

The authors declare that they have no known competing financial interests or personal relationships that could have appeared to influence the work reported in this paper.

**Acknowledgment:**

SGS, DET and GAS acknowledge the support of the Romanian Executive Agency for Higher Education, Research, Development and Innovation Funding (UEFISCDI) via grants RO-NO-2019-0601 MEDYCONAI and PN-III-P1-1.1-TE-2019-1339 OPTIGAN.

**Author contributions:**

The manuscript was written through contributions of all authors. All authors have given approval to the final version of the manuscript.